\scrollmode
\documentstyle[12pt]{article}
\catcode`\@=11

\@addtoreset{equation}{section}

\catcode`\@=11

\def\section{\@startsection{section}{1}{\z@}{3.5ex plus 1ex minus
   .2ex}{2.3ex plus .2ex}{\large\bf}}

\newskip\humongous \humongous=0pt plus 1000pt minus 1000pt

\newif\ifdtup

\def\thesection{\arabic{section}.}

\def\appendix{\setcounter{section}{0}
        \def\thesection{APPENDIX }}

\newcommand{\beq}{\begin{equation}}
\newcommand{\eeq}{\end{equation}}
\newcommand{\tg}{{\tilde g}}
\newcommand{\p}{{\hat p}}
\newcommand{\m}{{\hat m}}
\newcommand{\tp}{{\check p}}
\newcommand{\tchi}{{\tilde\chi}}
\newcommand{\ppi}{{\tilde\pi}}
\newcommand{\spin}{\left(\begin{array}{c}\tchi \\ \ppi\end{array}\right)}

\def\IR{{\hbox{{\rm I}\kern-.2em\hbox{\rm R}}}}
\def\IC{{\ \hbox{{\rm I}\kern-.6em\hbox{\bf C}}}}
\def\IZ{{\hbox{{\rm Z}\kern-.4em\hbox{\rm Z}}}}
\def\rref#1{(\ref{#1})}

\begin{document}

\begin{titlepage}
\vspace{.5in}
\begin{flushright}
UCD-91-16\\
August 1991\\
\end{flushright}
\vspace{.5in}
\begin{center}
{\Large\bf
(2+1)-Dimensional Chern-Simons Gravity\\as a Dirac Square Root\\}
\vspace{.4in}
{S.~C{\sc arlip}\footnote{\it email: Carlip@dirac.ucdavis.edu}\\
       {\small\it Department of Physics}\\
       {\small\it University of California}\\
       {\small\it Davis, CA 95616}\\{\small\it USA}}
\end{center}

\vspace{.5in}
\begin{center}
\large\bf Abstract
\end{center}

For simple enough spatial topologies, at least four approaches
to  $(2+1)$-dimensional quantum gravity have been proposed:
Wheeler-DeWitt quantization, canonical quantization in
Arnowitt-Deser-Misner (ADM) variables on reduced phase space,
Chern-Simons quantization, and quantization in terms of
Ashtekar-Rovelli-Smolin loop variables.  An important problem is to
understand the relationships among these approaches.  By explicitly
constructing the transformation between the Chern-Simons and ADM
Hilbert spaces, we show here that Chern-Simons quantization naturally
gives rise to spinorial wave functions on superspace, whose time
evolution is governed by a Dirac equation.  Chern-Simons quantum
gravity can therefore be interpreted as the Dirac square root of
the Wheeler-DeWitt equation.

\end{titlepage}

Over the past several years, it has become apparent that
$(2+1)$-dimensional Einstein gravity can serve as
a useful model for studying some of the
conceptual problems of quantum gravity in four spacetime
dimensions.  While attempts to quantize four-dimensional gravity have
thus far failed, several approaches have been worked out successfully,
in various degrees of detail, in three dimensions.  These include
Wheeler-DeWitt quantization \cite{Mart,HosNak,Visser}, quantization on
reduced phase space using Arnowitt-Deser-Misner (ADM) variables and
the York time slicing \cite{HosNak2,Mon}, and quantization in
terms of Chern-Simons variables and ISO(2,1) holonomies \cite{Wita,Cara}
or the closely related Ashtekar-Rovelli-Smolin loop variables
\cite{RovSmo}.  Three-dimensional gravity is not a terribly
realistic model --- among other shortcomings, it contains only finitely
many degrees of freedom --- but many of the basic issues of quantum
gravity do not depend on the dimension of spacetime, and even
a simple model may offer valuable insights.

The purpose of this paper is to relate the Chern-Simons quantization of
$(2+1)$-dimensional gravity to the quantum theory arising from the
Wheeler-DeWitt equation.  This work is in part a continuation
of reference \cite{Carb}.  In that paper, a start was made towards
comparing the fundamental operators in the Chern-Simons theory to those
of the ADM and Wheeler-DeWitt quantizations.  Here, that comparison is
extended to the Hilbert spaces and wave functions.

For spacetimes whose spatial topology is that of a
torus $T^2$, this comparison can be made completely explicitly, and,
as we shall see, the exact transformation between the two Hilbert
spaces can be worked out.
It is well known that the Wheeler-DeWitt equation is a
Klein-Gordon-like equation, and that models based on the true physical
degrees of freedom should be, in some sense, ``square roots.''
We shall see below that the ISO(2,1) quantum theory is a {\it Dirac}
square root, with wave functions that are most naturally understood
as spinors on superspace.

\section{Moduli and Holonomies}

We begin with a brief review of reference \cite{Carb}.  Let
$M$ be a spacetime with the topology $[0,1]\times \Sigma$, foliated
by surfaces $\Sigma_\tau$ of constant mean extrinsic curvature $\tau$.
The spatial metric $g_{ij}$ on a slice $\Sigma_\tau$ is conformal to
one of constant intrinsic curvature ($+1$ for spherical topology,
$0$ for the torus, $-1$ for higher genus):
\beq
g_{ij} = e^{2\lambda}\tg_{ij}\ .
\label{1x1}
\eeq
Moncrief \cite{Mon} has shown that the conformal factor
$\lambda$ is determined by the Hamiltonian constraint of ADM gravity.
The remaining momentum
constraints generate spatial diffeomorphisms of $\Sigma_\tau$, so the
physical degrees of freedom are constant curvature metrics $\tg$ on
$\Sigma$ modulo diffeomorphisms, along with their conjugate momenta.

The space of such constant curvature metrics is the moduli space
${\cal M}_\Sigma$ of $\Sigma$ \cite{Abikoff}.  In particular, for
$\Sigma = T^2$, flat metrics are parametrized by a complex modulus
$m=m_1+im_2$ with $m_2>0$.  ($m$ is usually called $\tau$ by
mathematicians, but we are using $\tau$ to denote the extrinsic
curvature.)  The modulus can be thought of as determining
a parallelogram with vertices $\{0,1,m,m+1\}$; a flat torus is
obtained by identifying opposite sides.  Not all values of $m$
correspond to distinct geometries, however.  The mapping class
group $\cal D$ --- the group of diffeomorphisms of $\Sigma_\tau$
not homotopic to the identity --- acts on the upper
half-plane $U$ with generators
\beq
S:m\rightarrow -{1\over m} \ , \qquad\quad T:m\rightarrow m+1\ .
\label{1x2}
\eeq
The true moduli space is the quotient space ${\cal M}_{T^2}=U/{\cal D}$,
and the physical phase space, obtained by adjoining
the conjugate momenta $p$, is the cotangent bundle $T^*{\cal M}$.

The Einstein action can now be rewritten in terms of $m$ and
$p$.  As Moncrief and Hosoya and Nakao show, we obtain
\beq
S = \int d\tau
\left( p_\alpha{\partial m^\alpha\over \partial\tau} - H \right) \ ,
\label{1x3}
\eeq
with the Hamiltonian
\beq
H = \tau^{-1}\left( h^{\alpha\beta}p_\alpha p_\beta \right)^{1/2} \ ,
\label{1x4}
\eeq
where $h_{\alpha\beta}$ is the constant negative curvature metric
on $U$, $h_{11}=h_{22}=m_2^{\ -2}$.  This action and Hamiltonian
are the starting points for ADM quantization.

In the Wheeler-DeWitt approach, rather than solving the Hamiltonian
constraint for $\lambda$, we impose it as a constraint on the
physical state space.  We therefore need a new variable corresponding
to $e^{2\lambda}$, which can be regarded as being conjugate to a
``time'' variable $T$.  As Hosoya and Nakao \cite{HosNak} have shown
(see also \cite{Visser}), there is a gauge choice in which only the
integral
\beq
v = \int_\Sigma d^2x \sqrt{\tg} e^{2\lambda}
\label{1x5}
\eeq
is relevant.  For general topologies, this gauge involves a choice of
time $T$ that differs from the mean extrinsic curvature.  For the
torus, however, $T = \tau$, and the Hamiltonian constraint takes
the form
\beq
{\cal H} = h^{\alpha\beta}p_\alpha p_\beta - v^2\tau^2 = 0 \ ,
\label{1x6}
\eeq
which must be imposed as an operator equation on the states.  Note
that $\cal H$ is essentially the square of the Hamiltonian \rref{1x4},
a first indication of the square root structure of the quantum theory.

In the ISO(2,1) holonomy approach, we
begin instead with the  first order form of the Einstein action.
As Witten has observed \cite{Wita}, the triad $e_\mu^{\ a}$ and the
spin connection $\omega_\mu^{\ ab}$ together constitute
an ISO(2,1) connection on $M$.  In terms of this connection, the
Einstein action is the standard Chern-Simons
action, and the field equations become the condition that the
connection be flat.

Now, a flat connection is determined by its holonomies (the
values of the Wilson loops), and the identity component of
the group of diffeomorphisms of $M$ acts on these holonomies by
conjugation.  A solution of the field equations is therefore
characterized by a group of ISO(2,1) holonomies --- i.e., a
representation of $\pi_1(M)$ in ISO(2,1) --- modulo conjugation.
Not all such holonomy groups can occur, however; it may be shown that
the SO(2,1) projection of any holonomy group arising from a solution
of the Einstein equations must be Fuchsian.  This condition picks
out a connected component in the space of holonomy groups, and
Mess \cite{Mess} has shown that any group lying in this component
corresponds to a solution of the field equations. (See also
\cite{Carc} for a description geared more towards physicists,
and \cite{Wael} for a closely related lattice approach).

The relationship between spacetimes and ISO(2,1) holonomies seems
rather abstract, but in fact one can construct a spacetime fairly
explicitly from a holonomy group.  Any subgroup $G$ of ISO(2,1)
acts on Minkowski space $V^{2,1}$ as a group of isometries.
If the SO(2,1) projection of $G$ is Fuchsian, this action is
properly discontinuous in some region ${\cal F}$ of $V^{2,1}$.
The quotient ${\cal F}/G$ is then a flat spacetime with topology
$[0,1]\times\Sigma$, and it (or some suitable
maximal extension) is the desired spacetime.

The space of ISO(2,1) holonomies has the structure of a cotangent
bundle, with a base space consisting of the SO(2,1) projections.
To obtain the set of physically distinct points in the phase space,
we must again divide out the mapping class group, which has
an action on the holonomies induced by its action
on $\pi_1(\Sigma)$.  It is not hard to show that in this representation,
the Hamiltonian is identically zero, a result that follows essentially
from the manifest diffeomorphism invariance of the holonomies
\cite{Mon2}.

For $\Sigma = T^2$, $\pi_1(M)$ is the Abelian group $\IZ\oplus\IZ$,
so a point in the physical phase space corresponds to a choice of
two commuting Poincar\'e transformations.  In the relevant
topological component of the space of holonomy groups, the
corresponding Lorentz transformations are parallel boosts,
which by overall conjugation can
be taken to be in the $x$ direction.  The holonomies are then
\begin{eqnarray}
  \Lambda_1&:& (t,x,y)\rightarrow(t\cosh\lambda+x\sinh\lambda,\,
           x\cosh\lambda+t\sinh\lambda,\,y+a) \nonumber \\
  \Lambda_2&:& (t,x,y)\rightarrow(t\cosh\mu+x\sinh\mu,\,
           x\cosh\mu+t\sinh\mu,\,y+b)\ \ .
\label{1x7}
\end{eqnarray}
The phase space is parameterized two pairs of canonically conjugate
variables $(a,\mu)$ and $(b,-\lambda)$, and it is not hard to show
that the action of the mapping class group corresponding to
\rref{1x2} is
\begin{eqnarray}
  S&:& (a,\lambda)\rightarrow(b,\mu),\quad
               (b,\mu)\rightarrow(-a,-\lambda)\nonumber \\
  T&:& (a,\lambda)\rightarrow(a,\lambda),\quad
               (b,\mu)\rightarrow(b+a,\mu+\lambda)\ \ .
\label{1x8}
\end{eqnarray}

The action of the group generated by $\Lambda_1$ and $\Lambda_2$ on
Minkowski space is worked out in detail in \cite{Carb}.  It is shown
that the quotient ${\cal F}/ \langle\Lambda_1,\Lambda_2\rangle$ is
indeed a spacetime with the topology $[0,1]\times T^2$, and that the
corresponding moduli and conjugate momenta of the slice $\Sigma_\tau$
of mean extrinsic curvature $\tau$ are
\beq
m =
\Biggl(a+{i\lambda\over\tau}\Biggr)^{-1}
\Biggl(b+{i\mu\over\tau}\Biggr) \ ,
\label{1x9}
\eeq
\beq
p = -i\tau\left(a-{i\lambda\over\tau}\right)^2 \ .
\label{1x10}
\eeq
In terms of the holonomies, the Hamiltonian \rref{1x4} is
\beq
H = {a\mu - \lambda b \over\tau} \ ,
\label{1x11}
\eeq
and it is not hard to check that the moduli and momenta \rref{1x9}
and \rref{1x10} obey Hamilton's equations of motion.

The relations \rref{1x9}--\rref{1x11} establish the classical
equivalence of the ADM and Chern-Simons formulations of
$(2+1)$-dimensional gravity on $M=[0,1]\times T^2$.  We now
turn to the quantum theories.

\section{Operators and Hilbert Spaces}

{}From now on, let us fix $M$ to have the topology
$[0,1]\times T^2$.  We begin with the simplest means of quantization,
the Wheeler-DeWitt method. In this approach, we impose the
Hamiltonian constraint \rref{1x6} on wave functions
$\psi(m,\bar m,\tau)$, with the operator substitutions
\beq
p_\alpha = -i{\partial\ \over\partial m^\alpha}\ ,
\qquad\quad v = -i{\partial\ \over\partial \tau} \ .
\label{2x1}
\eeq
With the standard operator ordering, the resulting Wheeler-DeWitt
equation is
\beq
{\cal H}\psi(m,\bar m,\tau) =
        \left[ \left(\tau{\partial\ \over\partial\tau}\right)^2
             + \Delta_0 \right]\psi(m,\bar m,\tau) = 0 \ ,
\label{2x2}
\eeq
where
\beq
\Delta_0 = -m_2^{\ 2}\left(\partial_1^{\ 2}+\partial_2^{\ 2}\right)
\label{2x3}
\eeq
is the scalar Laplacian on the upper half-plane with respect to the
Poincar\'e metric.  We should further require
that our wave functions $\psi$ be invariant under the modular
transformations \rref{1x2} --- that is, that they be automorphic
forms of weight zero --- and that they be square-integrable over a
fundamental region for the modular group.  The Laplacian restricted to
this domain is nonnegative, symmetric, and has a unique self-adjoint
extension.  Its spectrum has been studied by mathematicians,
and a fair amount is known about its eigenfunctions, which are
known as the weight zero Maass forms \cite{Iwan,Fay}.

The connection between Wheeler-DeWitt quantization and reduced phase
space ADM quantization is straightforward.  If we start with the
action \rref{1x3}, wave functions $\psi(m,\bar m,\tau)$ will again
be square-integrable automorphic forms of weight zero, with a time
evolution now governed by the Hamiltonian \rref{1x4}, i.e.,
\beq
\hat H = \tau^{-1}\Delta_0^{1/2} \ .
\label{2x4}
\eeq
The (positive) square root is defined by the spectral decomposition
of $\Delta_0$.  The Schr\"odinger equation in this approach is
simply
\beq
i{\partial\psi\over\partial\tau}(m,\bar m,\tau)
  = \hat H\psi(m,\bar m,\tau) \ ,
\label{2x4a}
\eeq
and any solution of \rref{2x4a} will clearly
also obey the Wheeler-DeWitt equation.  The ADM theory can thus
be viewed as the ordinary square root of the Wheeler-DeWitt equation.

The relationship between these quantizations and the Chern-Simons
formulation is more subtle, since the fundamental canonical variables
are different.  Our strategy will be to construct operators
representing the moduli $m(\tau)$ and their conjugate momenta from
the holonomies, and to reexpress the wave functions in terms of
eigenfunctions of these operators.  Note that from this point of view,
$\tau$ is simply a parameter labeling families of operators; its
physical significance as mean extrinsic curvature comes from the
correspondence principle.

The canonical variables in the Chern-Simons formulation on the
manifold $[0,1]\times T^2$ are $\{a,b,\lambda,\mu\}$.  The cotangent
bundle structure on the space of flat ISO(2,1) connections induces
a corresponding bundle structure on this parameter space, with
$\mu$ and $\lambda$ serving as coordinates on the base space (see,
for example, \cite{NelReg}).  We quantize the system by making the
substitutions
\beq
a = {i\over2} {\partial\ \over\partial\mu} \ ,
\qquad\quad b = -{i\over2} {\partial\ \over\partial\lambda} \ .
\label{2x5}
\eeq
Wave functions will now be functions $\chi(\mu,\lambda)$, invariant
under the modular transformations \rref{1x8} and square-integrable over
a suitable fundamental region.  The existence of such invariant
functions is not obvious, but they will be explicitly constructed below.

Our first task is to construct operators to represent the moduli
$m(\tau)$.  The classical correspondence \rref{1x9} allows us to do
so, up to questions of operator ordering. The appropriate ordering
is largely fixed by mapping class group invariance:
we must require that the transformations \rref{1x8} of the holonomies
induce the transformations \rref{1x2} of the moduli.  The ordering in
\rref{1x9} has been carefully chosen so that this is the case.
I do not know whether this choice (and, trivially, its transpose) are
unique in this respect, but if other such orderings exist, it is easy
to see that they must be quite complicated,
and in particular non-rational.

We therefore define moduli and momentum operators $\m$ and $\p$ by
replacing the holonomy variables in \rref{1x9}--\rref{1x10} with the
corresponding operators.  It is easy to check that $\m$ and $\p$
obey the correct canonical
commutation relations.  Note, however, that the momentum operators now
transform in a complicated way under the mapping class group:
\beq
S: \p\rightarrow \m^{\dagger 2}\p - i\m^\dagger \ , \qquad\quad
\p^\dagger\rightarrow \m^2\p^\dagger -3i\m\ .
\label{2x6}
\eeq
It is therefore inconsistent to represent these momenta as derivatives
with respect to the moduli.

Indeed, $\p$ and $\p^\dagger$ do not even transform as adjoints.
This, at least, can be cured by forming the combinations
$\tp = \p+\m_2^{\ -1}$ and $\tp^\dagger = \p^\dagger-\m_2^{\ -1}$,
which still obey the correct
canonical commutation relations, and which transform as
\beq
S: \tp \rightarrow \m^{\dagger 2}\tp + i\m^\dagger \ ,\qquad\quad
   \tp^\dagger \rightarrow \m^2\tp^\dagger -i\m\ .
\label{2x6a}
\eeq
The requirement that $\tp$ and $\tp^\dagger$ be adjoints then
determines the inner product
\beq
<\tchi_1|\tchi_2>=\int {d^2m\over m_2^{\ 2}} \,
                  {\bar\tchi_1}(m,\bar m)\tchi_2 (m,\bar m) \ ,
\label{2x7a}
\eeq
which is the expected Petersson inner product induced from the
Poincar\'e metric on the upper half-plane.

We can now represent $\tp$ and $\tp^\dagger$ as derivatives, or
equivalently set
\beq
\p = -2i{\partial\ \over\partial{\bar m}} - {1\over m_2} \ ,
\qquad\qquad \p^\dagger=-2i{\partial\ \over\partial{m}}+{1\over m_2}\ ,
\label{2x8}
\eeq
{\it provided} that these operators act on wave functions that
transform nontrivially under the mapping class group,
\beq
S: \tchi\rightarrow\left({m\over{\bar m}}\right)^{1/2}\tchi \ .
\label{2x9}
\eeq
The momenta can then be viewed as covariant derivatives; the phase
in \rref{2x9} accounts for the inhomogeneous terms in the
transformation \rref{2x6} and \rref{2x6a}.  Strictly speaking, to
be consistent with the relations among $S$ and $T$
viewed as elements of SL(2,$\IZ$) --- i.e., $S^2=(ST)^3=-I$ --- we
must include additional constant phases, and require that
\begin{eqnarray}
  S&:&
    \tchi\rightarrow e^{-{\pi i\over2}}
    \left({m\over{\bar m}}\right)^{1/2}\tchi\nonumber \\
  T&:&
    \tchi\rightarrow e^{\pi i\over6}\tchi\ \ .
\label{2x10}
\end{eqnarray}
This is precisely the requirement that the wave functions $\tchi$
transform as automorphic forms of weight $1/2$, that is, spinors on
moduli space \cite{Fay,Rankin}.

To check this behavior, we can examine the Hamiltonian \rref{1x11}.
It is easy to show that
\beq
\hat H = \tau^{-1}\left( m_2(\p^\dagger\p)^{1/2}
    - {1\over2}(\p^\dagger\p)^{-1/2}\p \right) \ ,
\label{2x11}
\eeq
or, inserting \rref{2x8},
\beq
{\hat H}^2 = \tau^{-2}\left(\Delta_0 +im_2\partial_1 - {1\over4}\right)
    = \tau^{-2}\Delta_{1\over2} \ ,
\label{2x12}
\eeq
where $\Delta_{1\over2}$ is the Maass Laplacian \cite{Fay} for
automorphic forms of weight $1/2$.  The holonomy formulation of
$(2+1)$-dimensional quantum gravity thus naturally gives rise to
wave functions that behave as spinors.

\section{Changing Representations}

The results of the previous section can be confirmed by another
approach. Inserting the operator representations \rref{2x5} directly
into the expression \rref{1x9} for the moduli, we can look for
the simultaneous eigenfunctions of $\m$ and $\m^\dagger$.
The eigenfunction with eigenvalues $m$ and $\bar m$ can be shown
to be
\beq
K(m,\bar m; \lambda,\mu,\tau)={\mu - m\lambda\over\pi m_2^{\ 1/2}\tau}
   \exp\left\{-{i\over m_2\tau}|\mu - m\lambda|^2\right\} \ .
\label{3x1}
\eeq
The $\mu$ and $\lambda$ dependence of $K$ is determined by the
eigenvalue equations, while the prefactor is fixed by the
normalization
\beq
\int d\mu d\lambda {\bar K}(m',\bar m';\mu,\lambda,\tau)
     K(m,\bar m;\mu,\lambda,\tau) = m_2^{\ 2}\delta^2(m-m') \ .
\label{3x2}
\eeq
(The factor of $m_2^{\ 2}$ on the right hand side of this expression
converts the ordinary delta function into a delta function with
respect to the Petersson inner product.)

An arbitrary wave function can now be expanded in terms of $K$,
\beq
\chi(\mu,\lambda) = \int {d^2m\over m_2^{\ 2}} \,
               K(m,\bar m;\lambda,\mu,\tau) \tchi(m,\bar m,\tau) \ .
\label{3x3}
\eeq
Note, however, that $K$ is not invariant under the simultaneous
mapping class group transformations \rref{1x2} and \rref{1x8}, but
rather transforms, up to constant phases, as an automorphic form
of weight $-1/2$.  For \rref{3x3} to be well-defined,
we must therefore once again require that $\tchi$ be an automorphic
form of weight $1/2$.  As promised, we then have
an explicit construction for wave functions $\chi(\mu,\lambda)$
invariant (up to constant phases) under modular
transformations of the holonomies.

The representations \rref{2x8} of the momenta can now be checked
explicitly.  For instance,
\begin{eqnarray}
\p\chi(\mu,\lambda)
 &=& -i\tau\left( {\hat a} - {i{\hat\lambda}\over\tau}\right)^2
   \chi(\mu,\lambda)
  = i\tau\left({1\over2}{\partial\ \over\partial\mu}
    - {\lambda\over\tau}\right)^2 \chi(\mu,\lambda) \nonumber \\
 &=& \int {d^2m\over m_2^{\ 2}}
    \left({3\over2m_2} - {i\over m_2^{\ 2}\tau}(\mu-m\lambda)^2\right)
    K(m,\bar m;\lambda,\mu,\tau) \tchi(m,\bar m,\tau) \nonumber \\
 &=& \int {d^2m\over m_2^{\ 2}}
    \left\{\left(2i{\partial\ \over\partial\bar m} + {1\over m_2}\right)
    K(m,\bar m;\lambda,\mu,\tau)\right\}\tchi(m,\bar m,\tau)\nonumber \\
 &=& \int {d^2m\over m_2^{\ 2}} K(m,\bar m;\lambda,\mu,\tau)
    \left(-2i{\partial\ \over\partial\bar m} - {1\over m_2}\right)
    \tchi(m,\bar m,\tau) \ ,
\label{3x4}
\end{eqnarray}
in agreement with \rref{2x8}.
Moreover, the time dependence of $\tchi(m,\bar m,\tau)$ can be determined
by the requirement that $\chi(\mu,\lambda)$ be $\tau$-independent.  For
the Hamiltonian given by \rref{1x11}, it is easy to check that
\beq
{\hat H}K(m,\bar m;\mu,\lambda,\tau)
   = -i{\partial\ \over\partial\tau}K(m,\bar m;\mu,\lambda,\tau) \ ,
\label{3x5}
\eeq
so
\begin{eqnarray}
0 = {\partial\chi\over\partial\tau}(\mu,\lambda)
&=&\int {d^2m\over m_2^{\ 2}}\left\{ i{\hat H}K(m,\bar m;\mu,\lambda,\tau)
   + K(m,\bar m;\mu,\lambda,\tau){\partial\ \over\partial\tau}\right\}
   \tchi(m,\bar m,\tau) \nonumber \\
&=&\int {d^2m\over m_2^{\ 2}} iK(m,\bar m;\mu,\lambda,\tau)
   \left\{ {\hat H}[\m,\p] - i{\partial\ \over\partial\tau} \right\}
   \tchi(m,\bar m,\tau) \ .
\label{3x6}
\end{eqnarray}
As expected, the time evolution of $\tchi(m,\bar m,\tau)$ is thus
generated by $\hat H$, now viewed as a functional of the moduli
and momenta $\m$ and $\p$.

\section{A Dirac Equation for Wave Functions}

The appearance of spinorial wave functions suggests that it should be
possible to express time evolution in the Chern-Simons model in terms
of a Dirac equation.  To see that this is so, first note that the
Laplacian $\Delta_{1\over2}$ can be written in terms of the Maass
operators
\cite{Fay}
\beq
K_n = (m-\bar m){\partial\ \over\partial m} + n \ , \qquad\quad
L_n = (\bar m-m){\partial\ \over\partial {\bar m}}-n = {K}^\dagger_{-n}
\label{4x1}
\eeq
as
\beq
\Delta_{1\over2} = -K_{-{1\over2}}L_{1\over2} \ .
\label{4x2}
\eeq
Here, $K_n$ maps forms of weight $n$ to forms of weight $n+1$, while
$L_n$ maps forms of weight $n$ to forms of weight $n-1$.
$K$ and $L$ have natural interpretations as chiral Dirac operators;
indeed, if we had demanded invariance under some Fuchsian group $G$
rather than the modular group, they would be the standard Dirac
operators on the Riemann surface $U/G$.

Now consider a wave function $\tchi(m,\bar m,\tau)$ that is an
(instantaneous) eigenfunction of the Hamiltonian \rref{2x11} with
nonvanishing eigenvalue $E$.  Since $\hat H$ is $\tau$-dependent, $E$
will be as well. Note, however, that $[\hat H(\tau),\hat H(\tau')]=0$,
so the Hamiltonian can be simultaneously diagonalized for all values
of $\tau$.  (This also means that no $\tau$-ordering operation is
needed in the definition of the evolution operator \cite{Isham}.)
Let us define a new automorphic form $\ppi$ of weight $-1/2$ by
\beq
L_{{1\over2}}\tchi(m,\bar m,\tau) = -\tau E\,\ppi(m,\bar m,\tau) \ .
\label{4x3}
\eeq
It is then easy to check that
\beq
\left( \begin{array}{cc} 0 & K_{-{1\over2}}\\
                         L_{1\over2} & 0 \end{array} \right) \spin
=\tau E \left(\begin{array}{cc} 1&0\\0&-1\end{array}\right) \spin \ ,
\label{4x4}
\eeq
which may be recognized as the time-independent Dirac equation.  Note
that the extra spin degree of freedom causes no difficulty in
interpretation: the wave function $\tchi$ is the unique weight $1/2$
(chiral) projection of a two-component spinor.

Like any Dirac equation, \rref{4x4} has both positive and negative
energy solutions.  Because of the simple $\tau$-dependence of the
Hamiltonian, however, the negative energy solutions have a particularly
easy interpretation.  $H$ is odd in $\tau$, so given an eigenfunction
$\tchi(\tau)$ with eigenvalue $E(\tau)$, the time-reversed wave
function $\tchi(-\tau)$ will have an eigenvalue $-E(\tau)$.
Classically, $\tau$ is the mean extrinsic curvature,
that is, the fractional rate of change of the volume of space
with respect to proper time.  Hence reversing the sign of
$\tau$, or $E$, amounts to replacing an expanding universe with a
contracting one.

If $\tchi$ is a zero-mode of $\hat H$, the definition \rref{4x3}
breaks down.  In that case, however,
\beq
0=\int{d^2m\over m_2^{\ 2}}\, \bar\tchi K_{-{1\over2}}L_{1\over2}\tchi
 =\int{d^2m\over m_2^{\ 2}}\, |L_{1\over2} \tchi|^2 \ ,
\label{4x5}
\eeq
so the Dirac equation \rref{4x4} holds with $\ppi=0$.  In fact, the
zero-mode can be computed explicitly: it is
\beq
\tchi_0(m,\bar m) = m_2^{\ 1/2}\eta^2(m) \ ,
\label{4x6}
\eeq
where $\eta(m)$ is the Dedekind eta function,
\beq
\eta(m) = e^{\pi im/12}\prod_{n=1}^{\infty} (1-e^{2\pi inm}) \ .
\label{4x7}
\eeq
This is an interesting wave function, representing a time-independent
toroidal universe with vanishing spatial volume but with a
nonzero expectation value for the modulus $m_2$.

\section{Conclusion}

We have seen that for the simplest nontrivial topology,
$M=[0,1]\times T^2$, Chern-Simons quantization of $(2+1)$-dimensional
gravity can be naturally understood as the Dirac square root of
Wheeler-DeWitt quantization.  This contrasts sharply with reduced phase
space ADM quantization, which is essentially an ordinary positive
square root.  This difference may help explain the comparative
simplicity of the Chern-Simons formulation.

This phenomenon depends on a choice of operator ordering, of course.
Our ordering was dictated by simplicity and modular invariance, but
it is possible that a more complicated choice --- in which the moduli
are not rational functions of the holonomy operators --- might
reproduce ADM quantization.  There seems to be no particular
justification for such a choice, however, except to match the
ADM theory.

An important question is whether these results are peculiar to
genus one.  For higher genus spaces, even the classical relationship
between moduli and ISO(2,1) holonomies becomes much more complicated,
and the connection between the corresponding quantum theories is quite
difficult to study.  One obvious conjecture can be made: Chern-Simons
wave functions for spaces of genus greater than one could be higher
dimensional automorphic forms on the corresponding Siegel upper
half-spaces.  It is possible that this conjecture
can be checked for the genus two case, for which the components of
the period matrix provide a good set of coordinates for moduli
space.

Of course, these simple $(2+1)$-dimensional results will not generalize
in any very straightforward manner to realistic $(3+1)$-dimensional
gravity.  The ``moduli space'' for a four-manifold, Wheeler's superspace,
is infinite dimensional, and the corresponding mapping class group
is likely to have complicated representations.  The $(2+1)$-dimensional
model does, however, provide a dramatic illustration of the
importance of the choice of canonical variables in the formulation
of quantum gravity, and it once again emphasizes the key role of the
mapping class group in quantization.

\vspace{2.5ex}
\begin{flushleft}
\large\bf Acknowledgements
\end{flushleft}

I would like to thank Vincent Moncrief, whose questions inspired this
paper, and the Aspen Center for Physics, which provided a productive
environment in which a portion of this work was completed.


\newpage
\begin{thebibliography}{99}

\bibitem{Mart} E.~Martinec, Phys.~Rev.~{\bf D30}, 1198 (1984).
\bibitem{HosNak} A.~Hosoya and K.~Nakao, Prog.~Theor.~Phys.~{\bf 84},
                 739 (1990).
\bibitem{Visser} M.~Visser, ``Wheeler-DeWitt Quantum Gravity in (2+1)
                 Dimensions,'' Washington University preprint, 1990.
\bibitem{HosNak2} A.~Hosoya and K.~Nakao, Class.~Quantum Grav.~{\bf 7},
                  163 (1990).
\bibitem{Mon} V.~Moncrief, J.~Math.~Phys.~{\bf 30}, 2907 (1989).
\bibitem{Wita} E.~Witten, Nucl.~Phys.~{\bf B311}, 46 (1988).
\bibitem{Cara} S.~Carlip, Nucl.~Phys.~{\bf B324}, 106 (1989).
\bibitem{RovSmo} A.~Ashtekar, V.~Husain, C.~Rovelli, J.~Samuel, and
                 L.~Smolin, Class.~Quantum Grav.~{\bf 6}, L185 (1989).
\bibitem{Carb} S.~Carlip, Phys.~Rev.~{\bf D42}, 2647 (1990).
\bibitem{Abikoff} W.~Abikoff, {\it The Real Analytic Theory of
                  Teichm\"uller Space} (Lecture Notes in Mathematics,
                  Vol.~820)(Springer, Berlin, 1980).
\bibitem{Mess} G.~Mess, Institut des Hautes Estudes Scientifiques
               preprint IHES/M/90/28 (1990).
\bibitem{Carc} S.~Carlip, Class.~Quantum Grav.~{\bf 8}, 5 (1991).
\bibitem{Wael} H.~Waelbroeck, Class.~Quantum Grav.~{\bf 7}, 751 (1990).
\bibitem{Mon2} V.~Moncrief, J.~Math.~Phys.~{\bf 31}, 2978 (1990).
\bibitem{Iwan} H.~Iwaniec, in {\it Modular Forms}, edited by
               R.~A.~Rankin (Ellis Horwood Ltd., Chichester, 1984).
\bibitem{Fay} J.~D.~Fay, J.~Reine~Angew.~Math.~{\bf 293}, 143 (1977).
\bibitem{NelReg} J.~E.~Nelson and T.~Regge, Nucl.~Phys.~{\bf B328},
                 190 (1989).
\bibitem{Rankin} R.~A.~Rankin, {\it Modular Forms and Functions}
                 (Cambridge University Press, Cambridge, 1977).
\bibitem{Isham} W.~F.~Blyth and C.~J.~Isham, Phys.~Rev.~{\bf D11},
                768 (1975).

\end{thebibliography}
\end{document}